\documentclass[]{spie}  
\usepackage[]{graphicx}

\title{The i-band Sky brightness and Transparency at Dome A, Antarctica} 

\author{Hu Zou\supit{a,b,c} and Xu Zhou\supit{a,c}
\skiplinehalf
\supit{a}National Astronomical Observatories, Chinese Academy of Sciences, Beijing 100012, China; \\
\supit{b}Graduate University of Chinese Academy of Sciences, Beijing 100049, China; \\
\supit{c}Chinese Center for Antarctic Astronomy, Purple Mountain Observatory, Chinese Academy of Sciences, Nanjing 210008, China
}

\authorinfo{Further author information: (Send correspondence to Dr. Xu Zhou)\\
Dr. Xu Zhou: E-mail: zhouxu@bao.ac.cn, Telephone: 86 010 6486 3043}

\begin{document} 
  \maketitle 

\begin{abstract}
Based on the observations of the Chinese Small Telescope ARray (CSTAR), the $i$ band 
observing conditions at Antarctic Dome A have been investigated. The over all variations of 
sky brightness and transparency are calculated and subsequently cloud cover, contributions to  
the sky background from various factors including aurorae are derived. The median sky brightness 
of moonless clear nights is about 20.5 mag arcsec$^{-2}$ in the SDSS $i$ band at the South 
Celestial Pole, which contains the diffused Galactic light of about 0.06 mag. There are no 
thick clouds in the year of 2008. Relatively strong aurorae are detected by their brightening 
the normal sky, which contribute up to about 2\% of the observed images.
\end{abstract}



\section{INTRODUCTION}
In site selections for ground-based optical and IR astronomy, some of the most important 
parameters are the night-sky brightness, seeing, atmospheric transparency, clear nights and 
humidity. The number of clear nights provides the usable observing time. The sky background, 
seeing and atmospheric transparency have effect on the observation quality and depth. Humidity 
affects the telescope and infrared transparency. 

The Antarctic plateau offers some attractive advantages for ground-based astronomical observations. 
It is a unique continent where there is no contamination from the artifacts. The average altitude in 
Antarctic plateau is more than 3000 m and the temperature is very low. Site testing over the past 
decade has revealed that Antarctica, relative to temperate latitude observatories, has lower infrared 
sky brightness, better free-atmosphere seeing, greater transparency, a lower turbulent boundary layer, 
and much lower water vapor content (see, e.g., reviews by Ref.~\citenum{ari05a,bur05,sto07}).

Dome A is located at the highest peak of the continent. It has an elevation of about 4093 m. The year-round 
average temperature is about $-50^\circ \mathrm{C}$, dropping to as low as $-80^\circ\mathrm{C}$ 
on occasion. Such high altitude, low temperature and special geographical position make it be very dry, 
cold and windless. It might be reasonably predicted that Dome A could be as good as or even a better 
astronomical site than Dome C (altitude of about 3250 m), with better seeing, higher transparency, and 
thinner surface layer. Ref.~\citenum{sau09} compared the sites Dome A, B, C, F and Ridge A and B in their 
cloud cover, free-atmosphere seeing, precipitable water vapor, temperature, and auroral emission, and 
concluded that, overall, Dome A might be the best of the existing bases for astronomical observations.

It is necessary that some facilities for site tests should be installed at Dome A to assess the site quality. 
In 2008 January, two Chinese astronomers deployed several instruments at Dome A including a small telescope array named 
CSTAR. With the observation data of 2008, we leaned the observing conditions such as the sky brightness, transparency 
and cloud covers. 

\section{Observations} 
CSTAR was built by the Nanjing Institute of Astronomical Optics \& Technology (NIAOT). It is composed of four 
co-aligned Schmidt telescopes on a fixed mounting pointing toward the South Celestial Pole. Each telescope has 
a different filter: $g$, $r$, $i$, and open. The detectors are Andor DV435 1k$\times$1k CCDs with 13 $\mu$m pixels, 
giving a plate scale of 15 arcsec pixel$^{-1}$. The entrance pupil diameter is 145 mm and the effective aperture is 
100 mm. Total field of view of each telescope is about 4.5 squared degrees. The main goals of this telescope are   
to detect variable stars and measure sky brightness, transparency, and cloud cover. Many telescope tests were performed 
at the Xinglong station before the telescope array was transported to Dome A \cite{zho10a}.

Due to defocus and other problems of control systems, the data from telescopes of $g$, $r$ and open bands are  
useless and only $i$ band data can be used to estimate those site parameters. In the whole winter, CSTAR 
observed over 310,000 images with integration times of 20 s or 30 s, giving a total exposure time of about 
1728 hr \cite{zho10b}. Data reductions for these CCD frames were also done by specifically designed processing pipelines. 

\section{Results}
\subsection{Sky Brightness}
There are sorts of sources contributing to the sky background such as scattering from
the Sun and the Moon, airglow, zodiacal light, aurorae, star light and interstellar dust scattering, 
extragalactic light, and artificial light contamination \cite{lei97,ben98a}. 
Artificial light pollution is essentially non-existent in Antarctica. Main contributions to the sky 
brightness come from the the atmospheric scattering of the light of the Sun and the
Moon. Zodiacal light might contribute up to half the intrinsic sky brightness if the Sun and Moon are 
down below the ground. For the pointing of CSTAR, the ecliptic longitude $\lambda$ of the South Celestial 
Pole is about $270^\circ$ and latitude $\beta$ is $-66.^\circ6$. Thus, the zodiacal light have little 
effect on the sky background. On the other hand, we expect an increased contribution from the Milky Way, 
since the Galactic latitude $b$ of the South Celestial Pole is $-27^\circ$, relatively close to
the Galactic plane. By the Pioneer 10 and 11 spacecraft observations of the total Galactic 
plus extragalactic sky background in blue (395--485 nm) and red (590--690 nm). We estimate the contribution 
of the diffused Galactic light to the sky brightness to be about 0.06 mag in $i$ band.

\begin{figure}
   \begin{center}
   \begin{tabular}{c}
   \includegraphics[height=7cm]{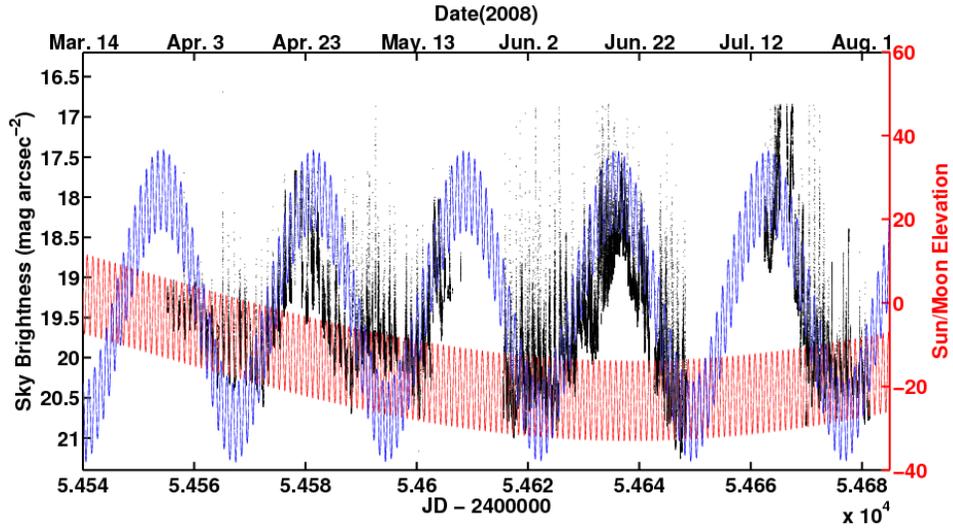}
   \end{tabular}
   \end{center}
   \caption{Sky brightness in $i$ band (black dots). The red and blue curves are the elevations of 
     the Sun and the Moon, respectively. \label{fig1} }
\end{figure}

\begin{figure}
   \begin{center}
   \begin{tabular}{cc}
   \includegraphics[width=0.5\textwidth]{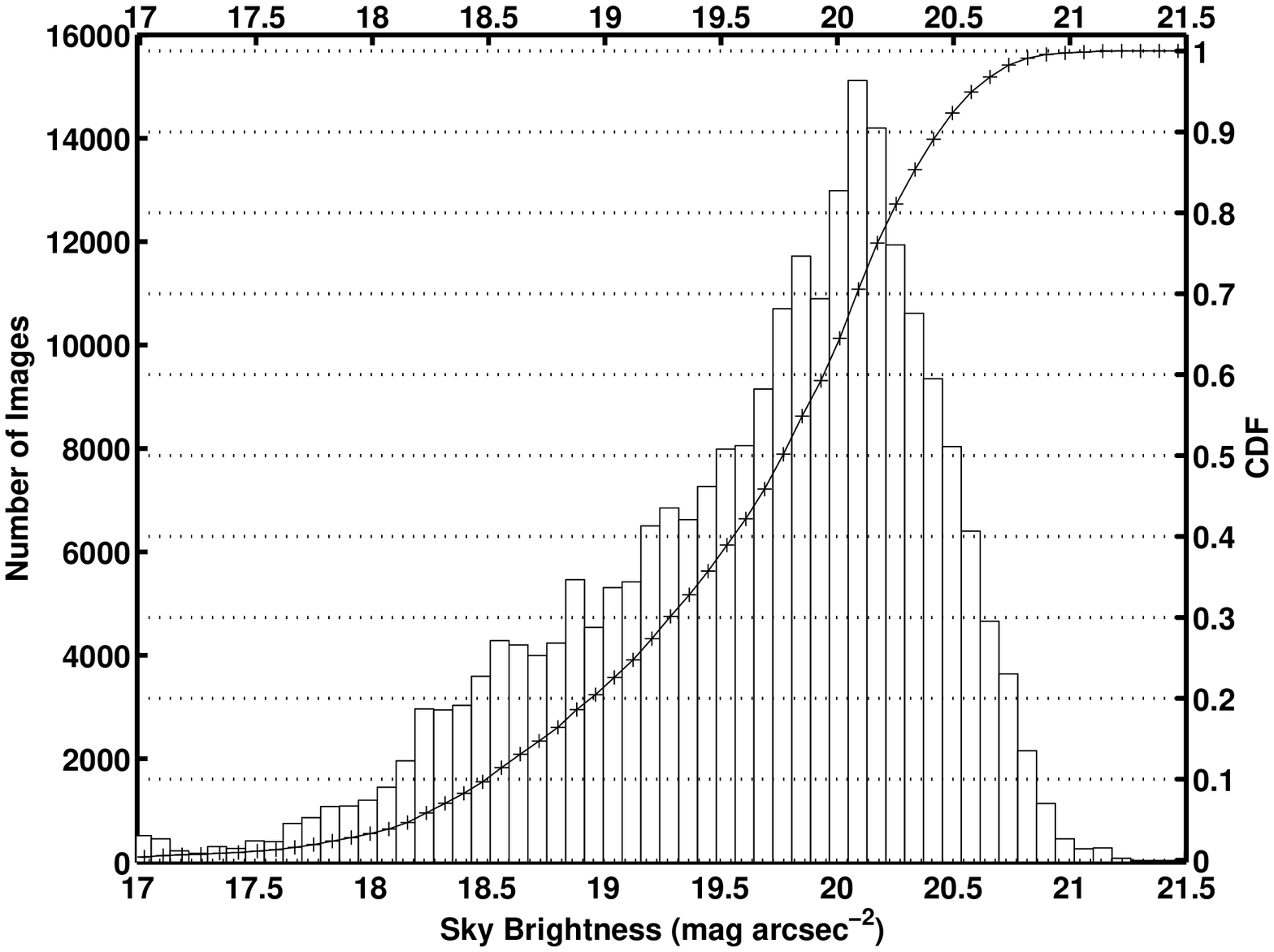}
   \includegraphics[width=0.45\textwidth]{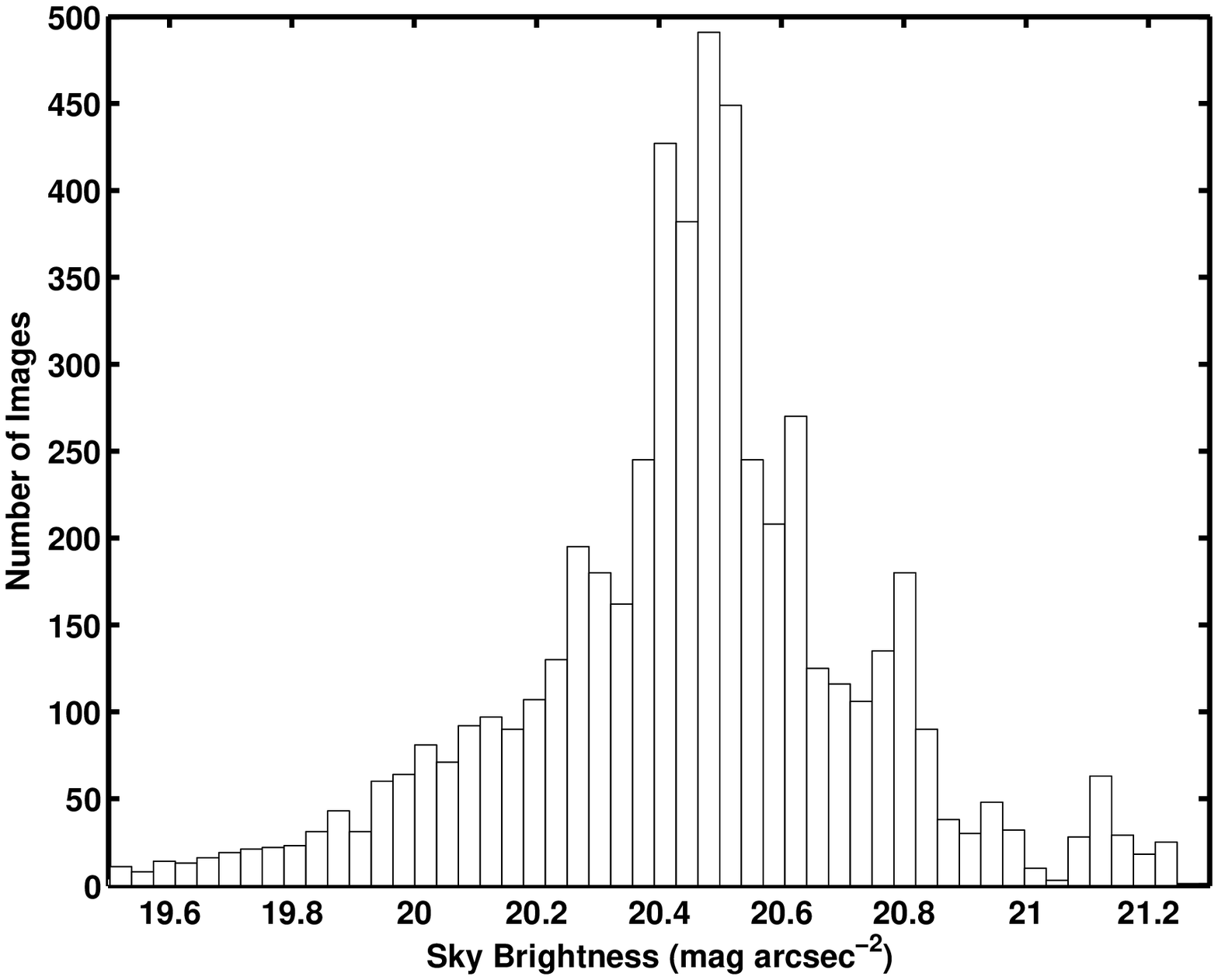}
   \end{tabular}
   \end{center}
   \caption{Left panel shows the histogram and the cumulative distribution function (CDF) of the $i$-band sky brightness
   distribution at Dome A during 2008. Right panel is the same information for the subset of
   images taken on moonless clear nights in 2008 June.\label{fig2}}
\end{figure}

Figure \ref{fig1} shows the sky brightness variations throughout the whole winter of 2008. 
At such far-southerly latitudes as Dome A, the Moon is always fairly full whenever it is 
above the horizon. Therefore, we can see that there is a strong correlation between lunar 
elevation (including the correction for parallax) and sky brightness. The situation should 
be seriously considered during the future astronomical observations at Dome A. Figure \ref{fig2} 
shows the sky brightness distribution at Dome A. The most probable sky brightness, across
all lunar phases in the left panel of the figure, is about 20.1 mag arcsec$^{-2}$ and the median 
is 19.8 mag arcsec$^{-2}$. The right panel of Figure \ref{fig2} shows the sky brightness 
distribution of images taken only on moonless and clear nights during 2008 June. In these images, 
the Sun and the Moon elevation is below $-18^\circ$ and there is no cloud. We find that 
the median sky brightness is 20.5 mag arcsec$^{-2}$. 

We compare our $i$-band sky brightness at Dome A with those of La Palma, Cerro Tololo,
and Paranal. The following median $i$-band sky brightnesses are calculated: 20.10
mag arcsec$^{-2}$ at La Palma (at sunspot minimum; Ref.~\citenum{ben98b}), 19.93 at Paranal
(at sunspot maximum; Ref.~\citenum{pat03}), 20.07 at Cerro Tololo (at sunspot minimum; Ref.~\citenum{wal87,wal88}) and
19.57 at Calar Alto (at sunspot maximum; Ref.~\citenum{san07}). The clear dark sky brightness
of 20.5 mag arcsec$^{-2}$ at Dome A can make a tentative conclusion that under moonless clear conditions, 
Dome A has a darker sky background than the above astronomical sites, even allowing for calibration and
other uncertainties of up to several tenths of a magnitude.

\subsection{Transparency}
Because CSTAR is fixed to point to the South Celestial Pole, we can not get the absolute 
atmospheric extinction. However, we can easily calculate the relative variation of the 
transparency by flux differences of selected bright stars between the observed frame and 
a reference frame which is chosen to be observed under good weather conditions. 

\begin{figure}
   \begin{center}
   \begin{tabular}{c}
   \includegraphics[height=7cm]{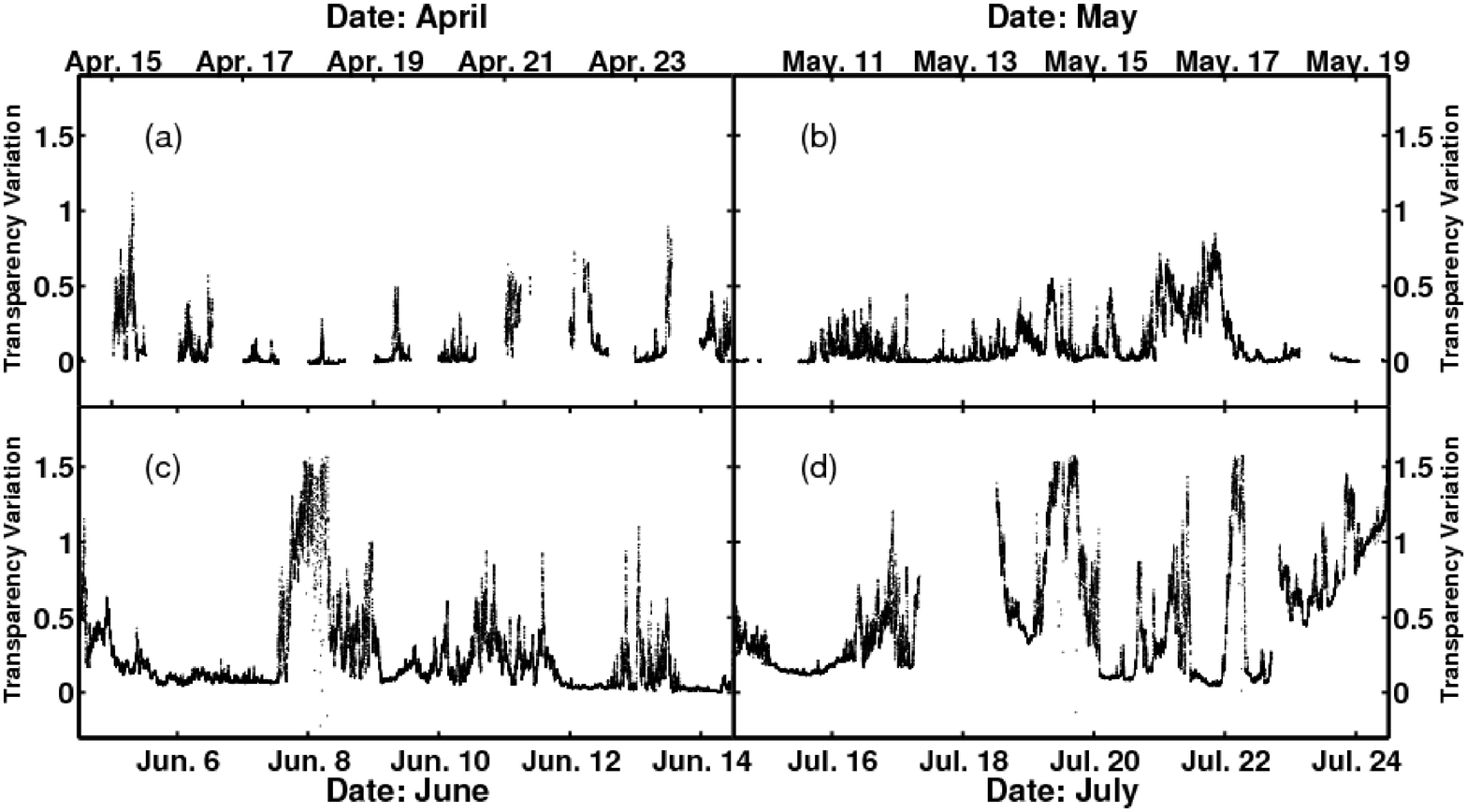} 
   \end{tabular}
   \end{center}
   \caption{(a) $i$-Band sky transparency variations (in magnitudes)
   from April 15 to April 25. The gaps are periods when sunlight made the sky too bright to
   observe. (b) May 10 to May 20, (c) June 5 to June 15, and (d) July 15 to July 25.\label{fig3}}
\end{figure}

Figure \ref{fig3} shows the relative transparency variation within 10 day periods in each 
month from April to July. From April 15 to June 2, the transparency was relatively stable 
with many days showing high transparency. In June and July, although the sky reaches its 
darkest, the transparency became a little worse. We use these observed relative transparency 
variations to infer the distribution of the optical thicknesses of cloud at Dome A during 2008. 
For a rough comparison with another observatory, Table \ref{tab3} shows the approximate fraction
of cloud of various thicknesses at Mauna Kea. The data for the table come from the Gemini
Observatory \footnote[1]{\it http://www.gemini.edu/sciops/telescopes-and-sites/observing-condition-constraints}. 
In this table, no thick cloud cloud cover at Dome A exists and it seems that there are more fraction of 
photometric conditions at Dome A than Mauna Kea. These conclusions are only tentative due to the limited 
observed sky region of CSTAR, and the fact that the data at Dome A only come from one winter of 2008.

\begin{table}
   \begin{center}
   \caption{The Comparison of Cloud Cover Between Mauna Kea and Dome A. \label{tab3}}
   \begin{tabular}{ccc|c}
   \hline \hline
   \multicolumn{3}{c|}{Mauna Kea (Gemini)} & Dome A\\
   \hline
   Cloud Cover & Extinction ($V$) & Fraction & Fraction \\
   \hline 
   Any other usable & $> 3$ & 10\% & 0 \\
   Cloudy & 2--3 & 20\% & 2\% \\
   Patchy cloud & 0.3--2 & 20\% & 31\% \\
   Photometric & $< 0.3$ & 50\% & 67\%  \\
   \hline
\end{tabular}
\begin{minipage}{\textwidth}
 Note: The definition of cloud cover is adopted from the Gemini Observatory. For comparison, 
we use $V - i = 0.07$ in extinction for the different tranparencies of these two bands as 
presented in the text. Note that the term `photometric' as used here is just one kind of cloud 
cover category and it is different from the normal term `photometric night'.
\end{minipage}
\end{center}
\end{table}

\subsection{Auroral Contribution} 
\begin{figure}
   \begin{center}
   \begin{tabular}{c}
   \includegraphics[height=7cm]{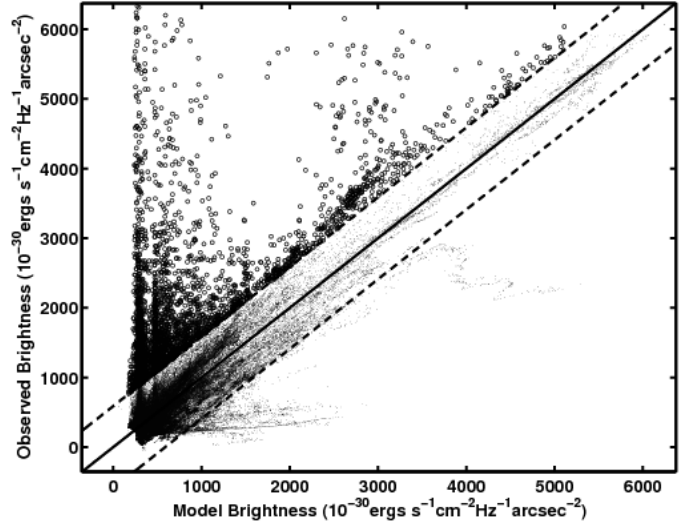} 
   \end{tabular}
   \end{center}
   \caption{Observed sky brightness corrected for the Sun and Moon contributions vs. the values
   predicted by the model from Equation (\ref{aurora}). The diagonal denotes where the observations 
   and model agree exactly. The two dashed lines are 3$\sigma$ above and below this level. The circles 
   show the observed data points, presumably affected by aurorae, with brightness brighter than 3$\sigma$ 
   above the model. The transparency variations are in magnitudes. \label{fig4}}
\end{figure}

Aurora is an important source polluting the observed images especially in the polar regions.
The generation of aurorae originates from the interaction between the high-speed charged particles from the 
Sun and the atoms and molecules in the upper atmosphere of the Earth under the force from the geomagnetic field.
Transition radiations mainly from the Nitrogen and Oxygen brighten the normal sky background. So we can 
detect the aurorae by removing the contributions to the sky brightness from the solar and lunar light and the extinction effect from 
the clouds. Equation \ref{aurora} presents our idea of auroral detection exactly. 
\begin{equation}
 F_{\textrm{sky}} = a (F_{\textrm{sun}}+F_{\textrm{moon}}) E + bE + c,\label{aurora}
\end{equation}
where $F_{\textrm{sky}}$ the sky brightness, $F_{\textrm{sun}}$ and $F_{\textrm{moon}}$ 
are the contributions from the Sun and Moon, $E$ is the extinction and a, b and c are the constants 
to be fitted. $F_{\textrm{sun}}$ and $F_{\textrm{moon}}$ can be modeled with the data which are dominated only by 
the contributions of either the Sun or the Moon. After fitting with the observed data, we can get the model sky brightness at 
any time. Then those frames which are 3$\sigma$ above the model brightness (as shown in circles in Figure \ref{fig4}) 
should be polluted by aurorae. We estimate that about 2\% of all the images are contaminated by aurorae.

\acknowledgments     
This study has been supported by the Chinese National Natural Science Foundation 
through grants 10873016, 10803007, 10473012, 10573020, 10633020, 10673012, and 10603006, 
and by the National Basic Research Program of China (973 Program), No.~2007CB815403.
This research is also supported by the Chinese PANDA International Polar Year project and 
the Polar Research Institute of China (PRIC). The support of the Australian Research Council 
and the Australian Antarctic Division for the PLATO observatory is gratefully acknowledged.
The authors thank all members of the 2008 and 2009 PRIC Dome A expeditions for their heroic 
effort in reaching the site and for providing invaluable assistance to the expedition astronomers
in setting up and servicing the PLATO observatory and its associated instrument suite. Iridium 
communications were provided by the US National Science Foundation and the United States Antarctic 
Program. Additional financial contributions have been made by the institutions involved in
this collaboration. We thank K.~Mattila for helpful discussions on the diffuse
Galactic light at the South Celestial Pole.



\begin{thebibliography}{1}
\bibitem{ari05a} Aristidi, E., et al., ``Site testing in summer at Dome C, Antarctica'', {\em A\&A}, {\bf 444}, pp.~651--659, 2005.
\bibitem{bur05} Burton, M.~G., et al., ``Science Programs for a 2-m Class Telescope at Dome C, Antarctica: PILOT, the Pathfinder for an International Large Optical Telescope'', {\em PASA}, {\bf 22}, pp.~199--235, 2005.
\bibitem{sto07} Storey, J.~W.~V., Lawrence, J.~S., \& Ashley, M.~C.~B., ``Site-testing in Antarctica'', {\em RevMexAA}, {\bf 31}, pp.~25--29, 2007.
\bibitem{sau09} Saunders, W., Lawrence, J. S., Storey, J. W. V., Ashley, M. C. B., Kato, S., Minnis, P., Winker, D. M., Liu, G., Kulesa, C., ``Where Is the Best Site on Earth? Domes A, B, C, and F, and Ridges A and B'',{\em pasp},{\bf 121}, pp.~976--992, 2009
\bibitem{zho10a} Zhou, X., et al.,``Testing and data reduction of the Chinese Small Telescope Array (CSTAR) for Dome A, Antarctica'', {\em RAA}, {\bf 10}, pp.~279--290, 2010
\bibitem{zho10b} Zhou, X., et al., ``The First Release of the CSTAR Point Source Catalog from Dome A, Antarctica'', {\em PASP}, {\bf 122}, pp.~347--353, 2010
\bibitem{lei97} Leinert, C., et al., ``1997 reference of diffuse night sky brightness (Leinert+ 1998)'', {\em VizieR Online Data Catalog}, {\bf 412}, 1997.
\bibitem{ben98a} Benn, C.~R., \& Ellison, S.~L., ``Brightness of the night sky over La Palma'', {\em New Astron. Rev.}, {\bf 42}, pp.~503--507, 1998.
\bibitem{ben98b} Benn, C.~R., \& Ellison, S.~L., ``La Palma Tech. Note, 115'', 1998.
\bibitem{pat03} Patat, F., ``UBVRI night sky brightness during sunspot maximum at ESO-Paranal'', {\em A\&A}, {\bf 400}, pp.~1183--1198, 2003.
\bibitem{wal87} Walker, A., ``N.O.A.O. Newsletter, No. 10'', 1987.
\bibitem{wal88} Walker, A., ``N.O.A.O. Newsletter, No. 13'', 1988
\bibitem{san07} S{\'a}nchez, S.~F., Aceituno, J., Thiele, U., P{\'e}rez-Ram{\'{\i}}rez, D., \& Alves, J.,``The Night Sky at the Calar Alto Observatory'', {\em PASP}, {\bf 119}, pp.~1186--1200, 2007.
\end{thebibliography}
\end{document}